\documentclass[journal]{IEEEtran}

\ifCLASSINFOpdf
\else
\usepackage[dvips]{graphicx}
\fi
\usepackage{url}
\usepackage{times}
\usepackage{babel}
\usepackage{multirow}
\usepackage{diagbox,amssymb}
\usepackage{slashbox}
\usepackage{newtxtext,newtxmath}
\usepackage{multirow}
\usepackage{amsfonts}
\usepackage{titling}
\usepackage{mathtools}
\usepackage{blindtext}
\usepackage{amsmath}
\usepackage{graphicx, epstopdf}
\usepackage{algorithm}
\usepackage{textcomp}
\usepackage[table]{xcolor}
\usepackage{caption}
\usepackage{subcaption}
\usepackage[square,numbers]{natbib}
\usepackage{algorithm}
\newcommand\Mycomb[2][^n]{\prescript{#1\mkern-0.5mu}{}C_{#2}}
\def\BibTeX{{\rm B\kern-.05em{\sc i\kern-.025em b}\kern-.08em
		T\kern-.1667em\lower.7ex\hbox{E}\kern-.125emX}}
\usepackage{natbib,etoolbox,lipsum,hyperref}

\newtheorem{theorem}{Theorem}[section]

\newtheorem{lemma}[theorem]{Lemma}

\begin{document}
	\title{Enhanced Fast Iterative Shrinkage Thresholding Algorithm For Linear Inverse Problem}
	%%%=========================%%%%
	\author{Avinash Kumar 
		%\thanks{This paragraph of the first footnote will contain the date on which you submitted your paper for review. It will also contain support information, including sponsor and financial support acknowledgment. For example, ``This work was supported in part by the U.S. Department of Commerce under Grant BS123456.'' }
		and Sujit Kumar Sahoo,  \IEEEmembership{Senior Member, IEEE}
		\thanks{School of Electrical Engineering, Indian Institute of Technology, Goa, India, sujit@iitgoa.ac.in.}
		%	\thanks{S. B. Author, Jr., was with Rice University, Houston, TX 77005 USA. He is now with the Department of Physics, Colorado State University, Fort Collins, CO 80523 USA (e-mail: author@lamar.colostate.edu).}
	}
	
	\markboth{Journal of \LaTeX\ Class Files, Vol. 14, No. 8, August 2015}
	{Shell \MakeLowercase{\textit{et al.}}: Bare Demo of IEEEtran.cls for IEEE Journals}
	%%%%%%%%%%%%=========================%%%%%%%%%%%%%

	\maketitle
	
	\begin{abstract}
 The linear inverse problem emerges from various real-world applications such as Image deblurring, inpainting, etc., which are still thrust research areas for image quality improvement. In this paper, we have introduced a new algorithm called the Enhanced fast iterative shrinkage thresholding algorithm (EFISTA) for linear inverse problems. This algorithm uses a weighted least square term and a scaled version of the regularization parameter to accelerate the objective function minimization. The image deblurring simulation results show that EFISTA has a superior execution speed, with an improved performance than its predecessors in terms of peak-signal-to-noise ratio (PSNR), particularly at a high noise level. With these motivating results, we can say that the proposed EFISTA can also be helpful for other linear inverse problems to improve the reconstruction speed and handle noise effectively.
	\end{abstract}
	

	\IEEEpeerreviewmaketitle

	\section{Introduction} \label{intro}
 We have taken Image deblurring as a linear inverse problem to demonstrate our proposed iterative reconstruction algorithm. A linear inverse problem in the form of a discrete system can mathematically be represented as,
	\begin{align}\label{eq1.1}
		b=Ax_{org}+w, 
	\end{align}
	where $b \in \mathbb{R}^M$, $A \in \mathbb{R}^{M \times N}$ are known, $w$ is the unknown noise, and $x_{org}\in \mathbb{R}^N$ is the "true" and unknown image to be estimated. Image deblurring is the process in which we estimate the underlying image from its blurred image. In the case of image deblurring problems, $b$ represents the blurred image, and $A$ represents the blur operator.
	
 A classical approach to this estimation problem is the Least Squares (LS) approach. In which we minimize the square error of the deblurred image for the problem (\ref{eq1.1}). It is represented as
	\begin{align}
		\label{eq1.2}
		\hat {x}_{LS} = \arg \min_{x}\frac{1}{2}\Vert Ax-b\Vert_2^2.
	\end{align}
	In many applications like image deblurring, it is often found that $A$ is ill-conditioned. Normally we have $M<N$, which indicates that the inverse problem is ill-poised and has infinitely many solutions. In such cases, the LS solution gives a huge norm and it becomes meaningless. To overcome this problem, we use different regularization methods to  stabilize the solution. Among them, the $l_1$ regularization method has received considerable attention in the image processing literature. 
 
 In regularization methods, the problem (\ref{eq1.1}) can be recast as 
	\begin{align}
		\label{eq1.3}
		\hat{x} =\arg \min_{x}f(x)+g(x),
	\end{align}
	where $f(x)=\frac{1}{2}\Vert Ax-b\Vert_2^2$, and $g(x)$ is the regularization function. $l_1$ regularization term refers to $\Vert x \Vert_1$, the sum of the absolute value of the components. Generally, it is used to induce sparsity in the optimal solution. In this case, $g(x)=\lambda\Vert x\Vert_1$, where $\lambda$ is the regularization parameter that trade-off $f(x)$ and $\Vert x \Vert_1$. To solve the problem (\ref{eq1.3}), there are a number of first-order gradient-based approaches like Iterative Shrinkage Thresholding Algorithm (ISTA) \cite{ISTA2004DAUB}, Fast ISTA (FISTA) \cite{beck2009fast}, Monotone FISTA (MFISTA) \cite{amir2009mfista}, and Improved FISTA (IFISTA) \cite{zulfiquar2015improved}.
 
	Starting with an initial guess $x_0$, a gradient descent (GD) based iterative solution of (\ref{eq1.2}) can be obtained by using the following step.
	\begin{align}
		\label{eq1.4}
		x_{k+1}=&x_{k}-\eta \nabla f(x_{k}),
	\end{align}
	where $x_{k}$ is $k^\text{th}$ estimation of $x$,  the step size is $\eta$, the gradient of $f(x)$ for $x_{k}$ is $\nabla f(x_{k})=A^T(Ax_{k}-b)$. The step size $\eta$ can be chosen in between $0$ and ${2}/{\lambda_{\text{max}}(A^{T}A)}$, with ${\lambda_{\text{max}}(A^{T}A)}$ being maximum eigenvalue of the matrix $A^TA$. In order to find an iterative solution to (\ref{eq1.3}), ISTA uses a proximal operator followed by the iteration mentioned in (\ref{eq1.4}). It iterates $x_{k+1}$ as
	\begin{align*}
		%\label{eq1.5}
		x_{k+1} =S_{\lambda\eta}[x_{k}-\eta\nabla f(x_{k})]
	\end{align*}
	where $ S_{\gamma} : \mathbb{R}^n \rightarrow \mathbb{R}^n$ is the proximal operator, defined as
	\begin{align*}
		%\label{eq1.6}
		S_{\gamma}[x]^{[i]}=\max\left(0, |x^{[i]}|-\gamma\right).\text{sign}\left(x^{[i]}\right),
	\end{align*}
 where $x^{[i]}$ refers to the $i^\text{th}$ entry of a vector $x$, $\gamma$ is a positive real number, and $\text{sign}\left(.\right)$ refers to the signum function that gives the sign of any input real number. This operator is also famously known as the shrinkage operator. 

Since ISTA iterations converge slowly at the rate of $O(\frac{1}{k})$, where $k$ is the iteration number. In order to improve the speed, a variant of ISTA called FISTA has been proposed by {Beck \textit{et al.}} FISTA integrates the Nesterove’s momentum into ISTA to achieve a faster convergence rate of $O(\frac{1}{k^2})$ \cite{beck2009fast}. The following are the iteration steps of FISTA.
	\begin{align*}
		%\label{eq1.7}
		x_{k+1} =&S_{\lambda\eta}[y_k-\eta \nabla f(y_k)]
	\end{align*}
where $y_{k}$ is the Nesterove’s momentum that combines previous two estimates $x_{k}$ and $x_{k-1}$. That is
	\begin{align}
	\label{NM}
	y_{k}=&x_{k}+\frac{\alpha_{k-1}-1}{\alpha_{k}}(x_{k}-x_{k-1}),
	\end{align}
where $\alpha_{k}$'s are generated sequence of numbers.

To further accelerate the convergence of FISTA, Zulfiquar \textit{et al.} has proposed IFISTA \cite{zulfiquar2015improved}. The contribution of IFISTA is to fast-forward the least square gradient descent iteration (\ref{eq1.4}) by $n$ steps as follows. 
\begin{align}
		\label{eq1.9}
		x_{k+1}=x_{k}-\eta W_n\nabla f(x_{k})
	\end{align}
	where $W_n$ is the weighting matrix and which is defined as
	\begin{align}
		\label{eq1.10}
		W_{n} =&\sum_{i=1}^{n}\Mycomb[n]{i} (-1)^{i-1}(\eta A^TA)^{i-1}.
	\end{align}
Since $W_n$ is pre-determined, the computation cost per iteration will remain the same as that of FISTA. The image de-blurring results in \cite{zulfiquar2015improved} show that the IFISTA algorithm provides competent image restoration capability with an improved convergence rate than the FISTA Algorithm.

We should note that the demonstration is only done for low Additive White Gaussian Noise (AWGN) scenarios. However, increasing the power of AWGN, the objective value starts diverging over the iteration of IFISTA, which can be seen from Fig. \ref{f1} of our result section.
In this paper, we have proposed an Enhancement of FISTA for linear inverse problems with high AWGN, and we evaluated its performance for the image deblurring problem. EFISTA accelerates the least square step of FISTA and effectively handles the noise incursion issue with a modified version of the regularization function. 
%EFISTA Algorithm works well for image-deblurring applications at high variance noise.

\subsection{Motivation}
Eq. (\ref{eq1.9}) can be seen as an $n^{th}$ order of exponential decay of the initial vector ${x}_0$. The same result can be achieved by $n$-iterations of (\ref{eq1.4}), which is only a single gradient descent step. Thus an improved convergence rate is achieved in IFISTA by replacing (\ref{eq1.4}) with (\ref{eq1.9}). This accelerated least square step works better for low AWGN. However, at high AWGN values, the noise incursion becomes prominent due to the accelerated LS. This can be noticed in the deblurring results of Fig. \ref{f13}. The motivation of this work is to retain the benefits of the accelerated least square while taking care of the noise incursion. As a result, we can Enhance the convergence of FISTA in presence of strong AWGN with an acceptable reconstruction quality. 
To answer this, we are going to update the regularization term in such a way that we can balance both the weighted least square term in the minimization problem.

\subsection{Contribution}
We have enhanced FISTA by accelerating the least square problem followed by a proper shrinkage to get a noise-free reconstruction. The proposed enhancement can also be extended to some modern algorithms like MFISTA-VA (MFISTA-variable acceleration) \cite{MFISTA-VA2019Zibetti}, IFISTA-BN \cite{wang2018novel}, OMFISTA (Over-relaxed MFISTA) \cite{OMFISTA2017Zibetti}.

\subsection{Organization} 
The development of EFISTA is described in section \ref{EFISTA}. We have shown the convergence of EFISTA in section \ref{Convergence}. In section \ref{Result}, Numerical results are presented and discussed, and finally, we draw some conclusions in section \ref{Conl}. 

%----------------------------------------------------------------------------------------------
\section{Enhanced FISTA} 
 \label{EFISTA}
Let us consider the $n$ steps of the gradient descent iteration mentioned in (\ref{eq1.4}) starting at an initial guess $x_0$.
\begin{align}
    \label{eq2.1}
    %x_1=&(I-\eta A^TA)x_{0}+\eta A^Tb,\nonumber\\
    %x_2=&(I-\eta A^TA)^2x_{0} +(I-\eta A^TA)\eta A^Tb + \eta A^Tb,\nonumber\\
    %\vdots \nonumber\\
    x_n=(I-\eta A^TA)^nx_{0} +\sum_{i=1}^{n}(I-\eta A^TA)^{i-1}\eta A^Tb,
\end{align}
By doing this we will get a closer estimate of the least square problem (\ref{eq1.2}) than a single step.
With an appropriate choice of $\eta$ the matrix $(I-\eta A^TA)$ will be a positive definite matrix. We can find a $W_n$ that satisfy the relation $(I-\eta A^TA)^n =(I-\eta W_n A^TA)$, and the Eq. (\ref{eq2.1}) can be written as follows.
\begin{align*}
 %   \label{eq5.2}
    x_n=&(I-\eta W_n A^TA)x_{0} +\eta W_n A^Tb,\nonumber\\
=&x_{0}-\eta W_n[A^T(Ax_{0} - b)]=x_{0}-\eta W_n\nabla f(x_{0})\nonumber
\end{align*}
The expression for $W_n$ can be derived using the properties of the Gram-matrix $A^TA$, as shown in Eq. (\ref{eq1.10}). Thus the IFISTA iteration step mentioned in Eq. (\ref{eq1.9}) is exactly identical to (\ref{eq2.1}), which is indeed an $n$ step fast forward of the gradient descent. The next step is to incorporate this accelerated least square into the regularized minimization problem Eq. (\ref{eq1.3}). 

Here the $l_1$ regularization used in \cite{beck2009fast} is considered. The solution to the problem in Eq. (\ref{eq1.3}) is obtained by applying the proximal operator $S_{\gamma}$ after the accelerated gradient descent step.
$$x_{k+1} =S_{\gamma}[x_{k}-\eta W_n\nabla f(x_{k})]$$ 
where $x_k$ is the previous estimate. The objective of the proximal operator is to perform denoising through shrinkage. The parameter $\gamma$ controls the strength of denoising happening to the estimated signal \cite{ppADMM}. Thus the $\gamma=\lambda\eta$ is set to perform appropriate denoising in the case of ISTA. 
In order to get a view of the noise power after the gradient descent step (\ref{eq1.4}), we can find the covariance matrix of $x_{k+1}$. 
\begin{align}
    \label{eq2.2}
C_{x_{k+1}} &= E[(x_{k+1}-Ex_{k+1})^T(x_{k+1}-Ex_{k+1})]\nonumber\\
&= E[A^Tww^TA]\eta^2 = A^TC_wA\eta^2
\end{align}
Here we have assumed the noise is uncorrelated with the previous estimates, and the additive noise vector is a zero mean white Gaussian noise. From the above expression, the noise variance at each pixel/index of the estimate $x_{k+1}$ can be upper bounded as
\begin{align*}
   % \label{nv1}
\sigma_{x}^2 \leq \lambda_{\text{max}}(A^{T}A)\sigma^2_w\eta^2,
\end{align*}
where $\sigma^2_w$ is the variance of the noise in the measurement, and $\lambda_{\text{max}}(A^{T}A)$ is the maximum eigenvalue of the matrix $A^TA$.

Using a similar analysis as Eq. (\ref{eq2.2}), in the scenario of the accelerated gradient descent step as mentioned in (\ref{eq1.9}), the noise variance in the estimate will be as follows,
$$C_{x_{k+1}} = W_nA^TC_wAW_n^T.$$
Thus the noise variance at each pixel/index of the estimate $x_{k+1}$ can be upper bounded as
\begin{align*}
   % \label{nv2}
\sigma_{x}^2 \leq \lambda_{\text{max}}(A^{T}A)\lambda_{\text{max}}(W_nW_n^T)\sigma^2_w\eta^2.
\end{align*}
We can notice that the maximum noise variance got scaled by the maximum eigenvalue of the matrix $W_nW_n^T$ in the case of accelerated gradient descent. Since $W_n=W_n^T$ is a symmetric matrix, we can write $\lambda_{\text{max}}(W_nW_n^T)=\lambda^2_{\text{max}}(W_n)$. Taking the scaling of the noise variance into account better denoising can be performed by scaling the shrinkage threshold parameter $\gamma=\lambda\eta$. Thus we proposed to use a new threshold parameter as $\gamma=p\lambda\eta$ to perform proper denoising.  
$$x_{k+1} =S_{p\lambda\eta}[x_{k}-\eta W_n\nabla f(x_k)]$$ 
where $p \in \left[1, \lambda_{\text{max}}(W_n)\right]$, as the factor of increment in the noise standard deviation. Similar to FISTA, the accelerated gradient descent can be used on Nesterove's momentum mentioned in Eq. (\ref{NM}). The following is the complete description of the algorithm.  
\begin{algorithm}[H]
\caption{: Enhanced FISTA (EFISTA)}\label{alg:three}
\textbf{Input}: $b$, $A$, $\lambda \propto \sigma_w$, {$\eta\in \left[0, {\lambda^{-1}_{\text{max}}(A^{T}A)}\right]$}, ${p \in \left[1, \lambda_{\text{max}}(W_n)\right]}$\\
%\KwInitialization{Write here the expected result},
\textbf{Initialization}: ${W_{n} =\sum\limits_{i=1}^{n}\Mycomb[n]{i} (-1)^{i-1}(\eta A^TA)^{i-1}}$, $k=0$,\\ $\alpha_0 =1$, $x_0$, and $y_0 =x_0$\\
\textbf{While}{ stopping criteria is not met,}{
\begin{align*}
x_{k+1} =& S_{p\lambda\eta}[y_{k}-\eta W_n\nabla f(y_k)]\\
\alpha_{k+1} =&\frac{1+\sqrt{1+{{4\alpha_k}^2}}}{2}\\
y_{k+1}=&x_{k+1}+\frac{\alpha_k-1}{\alpha_{k+1}}(x_{k+1}-x_{k})\\
k=&k+1
\end{align*}
}
\textbf{Output}:\;$\hat{x} = x_N$
\end{algorithm}

%-------------------------------------------------------------------------------------------------------------------------------------------
\section{Convergence analysis of the proposed EFISTA algorithm}
\label{Convergence}
	Let us consider the regularized minimization problem given in (\ref{eq1.2}), where the objective function
	\begin{align*}	%\label{MLM}
		 F(x) = f(x)+ g(x).
	\end{align*}
The convergence rate of the EFISTA is obtained in a similar manner to FISTA \cite{beck2009fast}. First of all, we approximate $F(x)$ at a given point $z$ as 
	\begin{align}
		\label{eq13}
		Q(x,z) = &f(z)+(x-z)^{T}\nabla{f(z)}\nonumber\\
		& +\frac{1}{2\eta}\Vert x - z\Vert_{W_n^{-1}}^2 + pg(x)
	\end{align}
	where $Q(x,z)\geq F(x)$ for a choice of $0<\eta<\infty$, $W_n^{-1} > 0$, and $p \in \left[1, \lambda_{\text{max}}(W_n)\right]$. The next step is to minimize this above approximation function $Q(x,z)$. After ignoring constant terms, we can get the solution as
	\begin{align*} 	
	%\label{eq15}
		x=S_{p\lambda\eta}[z-\eta W_n \nabla f(z)].
	\end{align*}
	By  putting $x=x_{k+1}$, and $z=y_{k}$, we obtain EFISTA update step, 
	\begin{align} 
	\label{eq16}
		x_{k+1}=S_{\gamma}[y_{k}-\eta W_n \nabla{f(y_{k})}],
	\end{align}
	where $\gamma=\eta \lambda p$ is threshold parameter. Eq. (\ref{eq16}) can be seen as a sequence generating function in the EFISTA iterations, ${x_{k+1}=P_{\gamma}(y_{k})}$.
 
	In the next step, we extend Lemma 2.3 given in \cite{beck2009fast} for the function $F(x)$ and $Q(x,z)$ for considering $n=1$, and $W_n=I$.
	%-----------------------------------
	\begin{lemma}\label{lemma1}
    For any $x$ , the relation
	\begin{align} 	
		F(x) - F(P_{\gamma}(z)) \geq &\frac{1}{2\eta}\Vert P_{\gamma}(z)-z \Vert_{W_n^{-1}}^2\nonumber\\
		&+ \frac{1}{\eta}(z-x)^TW_n^{-1}(P_{\gamma}(z)-z)
	\end{align}
	holds true.
	\end{lemma}
	%------------------------------------
 The proof of this lemma will be similar to \cite{zulfiquar2015improved}. Similarly, we can extend Lemma 4.1 given in \cite{beck2009fast} as follows.
%-------------------------------------
	\begin{lemma}\label{lemma2}
	The sequence $(x_k, y_k)$ generated by EFISTA satisfies 
	$$2\eta(\alpha_{k-1}^2d_k-\alpha_{k}^2d_{k+1})\geq \Vert u_{k} \Vert_{W_n^{-1}}^2-\Vert u_{k-1} \Vert_{W_n^{-1}}^2$$
	where,\; $d_k=F(x_k)-F(x^*)$  and $u_k=\alpha_k x_{k+1}-(\alpha_k-1)x_{k}-x^*$.
\end{lemma}	
%--------------------------------
Here, we can interpret $d_k$ as the difference of function value at $x_k$ w.r.t functional value at optimal $x^*$.

With the help of the above two Lemmas, we can prove The following lemma \cite{zulfiquar2015improved}. 
%------------------------------------------------------------
\begin{lemma}\label{lemma3}
Let $(x_k, y_k)$ be the sequence generated by the EFISTA algorithm. Then for any $k >1$, we obtain 
	$$F(x_k) -F(x^*) \leq \frac{\Vert x_0-x^*\Vert_{W_n^{-1}}}{(k+1) ^2}$$
\end{lemma}
This is the main result of the convergence of EFISTA, which shows that EFISTA minimizes the difference in the objective function value between the optimal solution and the initial guess at a rate proportional to $1/(k+1)^2$ in $k$ iterations. 
%-----------------
%\begin{lemma}\cite{beck2009fast}
%The positive sequence $\{\alpha_k\}$ generated in FISTA via $\alpha_{k+1} %=\frac{1+\sqrt{1+{{4\alpha_k}^2}}}{2}$ with $\alpha_{0}=1$ satisfies %$\alpha_k\geq \frac{k+1}{2}$ for all $k\geq 1$
%\end{lemma}
%---------------------------------------------------------------------------------------------------------------------------------------------------------------------------
\section{Numerical result}\label{Result}
In this section, we compare the performance of EFISTA with IFISTA and FISTA for image deblurring applications. The original image in all the test cases is normalized to $(0,1)$. A gaussian blur of size [$7,7$] with a standard deviation of $4$ is used to blur the original image. The blurred images are contaminated with AWGN of zero-mean and standard deviation $\sigma$. In all the tests, we have used the reflexive boundary conditions \cite{hansen2006matlabcode} and the shrinkage operation is performed on an 8-stage CDF 9/7 wavelet transform domain. We have used a weighting matrix $W_8$ as described in Eq. (\ref{eq1.10}) if it is not stated. The regularization parameter is set to $\lambda =10\sigma^2$, the step size $\eta=1$. All tests are carried out using MATLAB R$2021$b on an Intel Core i$7-8700$ CPU $3.20$GHz processor.
	\begin{figure}[th]
		\centering
\includegraphics[height=4.5cm,trim={0.3cm 0.1cm 1.0cm 0.6cm},clip]{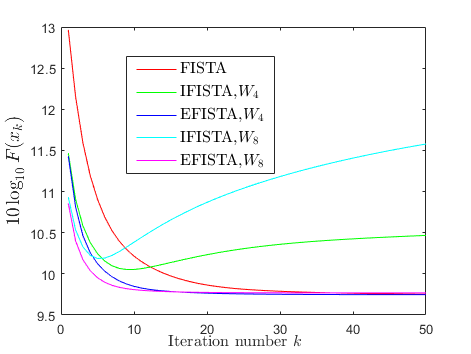}
		\caption{objective function vs iteration number}
		\label{f1}
	\end{figure}
 
In Test $1$, we have used the blurred Cameraman image of size $256 \times 256$ with $\sigma=10^{-2}$. We have calculated the weighting matrices $W_8$ and $W_4$ as described in Eq. (\ref{eq1.10}), which can easily be done using spectral decomposition. We have considered $p$ value as $8=\lambda_{max}(W_8)$ for $W_8$ and $4=\lambda_{max}(W_4)$ for $W_4$. We have iterated the above algorithms for $50$ number of iterations and have taken $10$ number of trials to plot the curve between the objective function value $F(x_k)$ and iteration number $k$ in Fig. \ref{f1}. In this figure, we can see that the objective function value for both EFISTA and IFISTA converges faster than FISTA. However, IFISTA starts diverging after the initial few iterations for both the weighting matrices due to noise incursion. We can notice that EFISTA starts converging after $35^{th}$ iteration and $15^{th}$ iteration for $W_4$ and $W_8$ matrices, respectively.
	\begin{figure}[h]
		\centering
	\includegraphics[height=6cm,trim={0.3cm 0.1cm 1.0cm 0.7cm},clip]{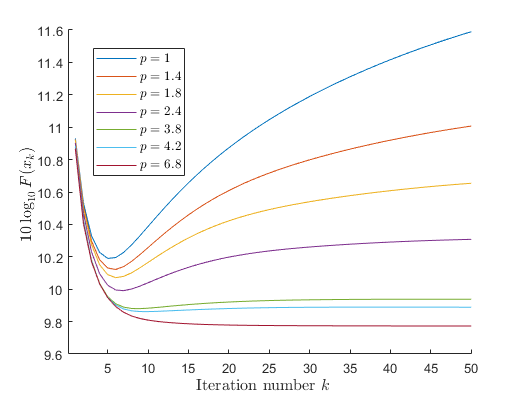}
		\caption{Objective function vs iteration number at $W_8$}
		\label{f3}
	\end{figure}
 	\begin{figure}[tbh]
		\centering
	\includegraphics[scale=.6,trim={0.45cm 0.15cm 1.0cm 0.6cm},clip]{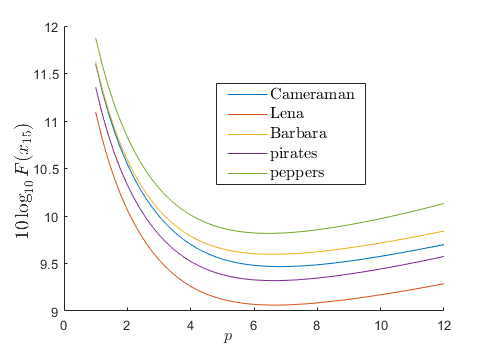}
		\caption{Objective function value at $15^{th}$ iteration vs $p$  at  $W_8$ for different test image}
		\label{f5}
	\end{figure}
 
In Test $2$, we repeat Test $1$ only for EFISTA with weighting matrix $W_8$ for different values of $p$ and plotted in Fig. \ref{f3}. From this test, we observed that there is no more divergence at $p=6.8$. In Test $3$, we repeat Test $2$ of EFISTA with the same parameters for different standard test images of size $256 \times 256$. The objective function value at $15^{\text{th}}$ iteration vs $p$ values is plotted  in Fig. \ref{f5}, which shows a clear minimization of the objective function when $p$ is closer to $\lambda_{max}(W_8)$. 
\begin{figure}[tbh]
 \centering
 \begin{subfigure}[b]{0.24\textwidth}
     \centering
     \includegraphics[width=\textwidth]{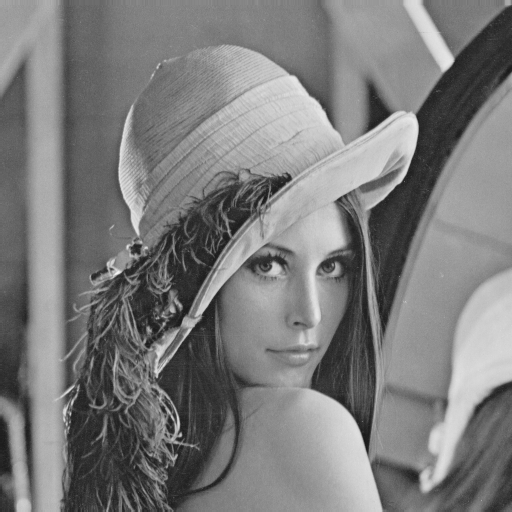}
     \caption{Original }
        \label{f11}
 \end{subfigure}
 \hfill
 \begin{subfigure}[b]{0.24\textwidth}
     \centering
     \includegraphics[width=\textwidth]{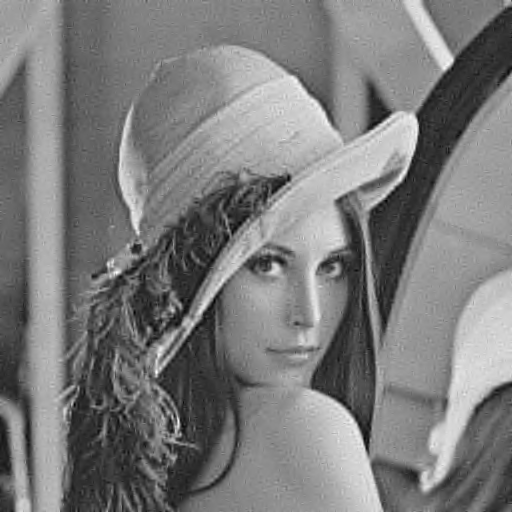}
     \caption{FISTA}
        \label{f12}
 \end{subfigure}
 \hfill
 \begin{subfigure}[b]{0.24\textwidth}
     \centering
     \includegraphics[width=\textwidth]{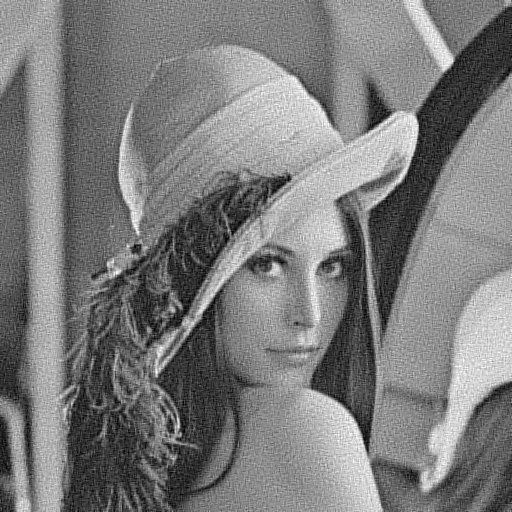}
     \caption{IFISTA}
       \label{f13}
 \end{subfigure}
 \begin{subfigure}[b]{0.24\textwidth}
     \centering
     \includegraphics[width=\textwidth]{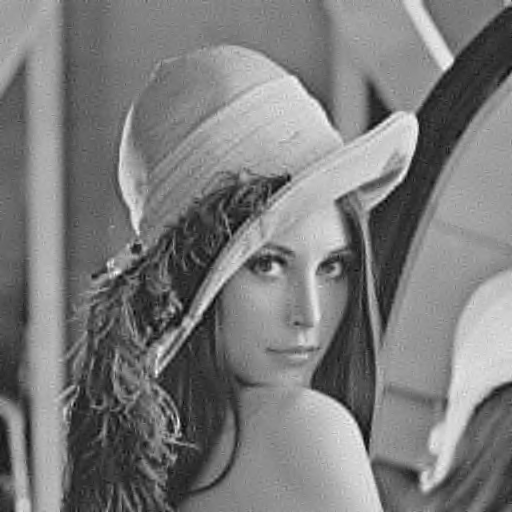}
     \caption{EFISTA}
        \label{f14}
 \end{subfigure}
 \hfill
    \caption{De-blurrred images for Lena test image}
    \label{f15}
\end{figure}

%\begingroup
\setlength{\tabcolsep}{3pt} % Default value: 6pt
\begin{table}[tbh]
	\centering
	\begin{tabular}{|c|c|c|c|c|c|c|}
		\hline
Noise level&	Algorithms & \textcolor{blue}{Cameraman} & Lena & Barbara & Pirate & \textcolor{blue}{Peppers}\\
		\hline
	\multirow{3}{4em}{\centerline{$\sigma=10^{-2}$}}& FISTA&  25.39 &  30.21  & 24.34 &  28.30  &\bf 27.50 \\
	&	IFISTA&  23.59 &  25.26 &  22.88&   24.80&   24.49\\
	&	 EFISTA &\bf 25.39 &\bf 30.28 &\bf  24.34 &\bf 28.32 &  27.49 \\\hline
		\hline
	\multirow{3}{4em}{\centerline{$\sigma=10^{-3}$}}& FISTA& 30.05  & 33.47  & \bf 27.78  & 32.17   &32.07 \\
	&	IFISTA& 28.79  &31.06  & 27.24  & 30.43  &30.13\\
	&	 EFISTA& \bf 30.05  &\bf 33.56 &  27.77 &  \bf 32.24  & \bf 32.13 \\\hline
	\end{tabular}
	\caption{Average PSNR Value for different noise levels}
	\label{tab1}
\end{table}
%\endgroup\\

In Test $4$, we have done deblurring of five standard test images at two noise levels, i.e., $\sigma=10^{-2}$ and $\sigma=10^{-3}$. From Test $1$ it is evident that EFISTA and IFISTA converge faster than FISTA. Therefore, in each image reconstruction, if we have used $K$ iterations for FISTA, then we have used $K/3$ iterations for both IFISTA and EFISTA. We used $K=45$ iteration for $\sigma=10^{-2}$, $K=180$ iteration for $\sigma=10^{-3}$. The PSNR values of the reconstructed images for all algorithms are presented in Table. \ref{tab1}. We have also plotted deblurred images for noise level, $\sigma=10^{-2}$ in Fig. \ref{f15}. We can notice that EFISTA performs a better reconstruction, while IFISTA incurred noise.
%We have noted the execution time for Test $4$, which is mentioned in Table \ref{tab2}. 

%-----------------------------
\section{Conclusion}\label{Conl}
An Enhancement Fast Iterative Shrinkage Thresholding Algorithm (EFISTA) is developed in this paper by introducing a scaled regularization term with the weighted least square term in the minimization problem. The EFISTA regulates the noise incursion effectively while taking advantage of the accelerated convergence and provides competitive image restoration performance. The simulation results confirm its superior execution speed and reconstruction quality. In the case of high AWGN, the simulation shows a better PSNR for EFISTA over IFISTA. The EFISTA Algorithm could be used as an alternative to its predecessor. 
	%\newpage
	%\renewcommand\refname{\vskip -1cm}
	%\begin{flushleft}
	%\section*{\large {References}}
	%\bibliographystyle{harvard}
	%\bibliography{sample.bib}
	%\end{flushleft}

	%\section{References}
	%\medskip
	%\medskip
	\bibliographystyle{ieeetr}
	\bibliography{Referencepapers}

\end{document}